\documentclass[a4paper,11pt]{article}

\usepackage[a4paper,left=80pt,right=80pt,top=77pt,bottom=100pt]{geometry}
\usepackage[latin1]{inputenc}
\usepackage{dsfont}
\usepackage{amsfonts,amsmath,amssymb}
\usepackage{amsthm}
\usepackage{graphicx}
\usepackage[small,bf]{caption}
\usepackage{subcaption}
\usepackage{booktabs}
\usepackage[numbers, sort]{natbib}
\allowdisplaybreaks[1]
\usepackage{color}
\usepackage{ulem}
\usepackage{tikz}
\usepackage{multirow}
\usetikzlibrary{arrows}

\usepackage[pagebackref=false]{hyperref}

\def\bal#1\eal{\begin{align}#1\end{align}}
\newcommand{\be}{\begin{equation}}
\newcommand{\ee}{\end{equation}}
\newcommand{\besub}{\begin{subequations}}
\newcommand{\eesub}{\end{subequations}}
\newcommand{\ba}{\begin{array}}
\newcommand{\ea}{\end{array}}
\newcommand{\bi}{\begin{itemize}}
\newcommand{\ei}{\end{itemize}}
\newcommand{\nn}{\nonumber}

\newcommand{\vev}[1]{\ensuremath{\langle #1 \rangle}}
\newcommand{\eV}{{\rm eV}}
\newcommand{\GeV}{{\rm GeV}}
\newcommand{\TeV}{{\rm TeV}}

\newcommand{\ckm}{{\text{\tiny CKM}}}
\newcommand{\mns}{{\text{\tiny PMNS}}}

\def\nuL{\nu}

\def\nl{.}
\def\nlone{.}

\begin{document}

\begin{titlepage}

\begin{center}
{\LARGE
Inverse neutrino mass hierarchy in a flavour GUT model
}
\\[15mm]

Stefan Antusch$^{\star\dagger}$\footnote{Email: \texttt{stefan.antusch@unibas.ch}},~
Christian Gross$^{\star}$\footnote{Email: \texttt{christian.gross@unibas.ch}},~
Vinzenz Maurer$^{\star}$\footnote{Email: \texttt{vinzenz.maurer@unibas.ch}},~
Constantin Sluka$^{\star}$\footnote{Email: \texttt{constantin.sluka@unibas.ch}}
\end{center}
\addtocounter{footnote}{-4}

\vspace*{0.20cm}

\centerline{$^{\star}$ \it
Department of Physics, University of Basel,}
\centerline{\it
Klingelbergstr.\ 82, CH-4056 Basel, Switzerland}

\vspace*{0.4cm}

\centerline{$^{\dagger}$ \it
Max-Planck-Institut f\"ur Physik (Werner-Heisenberg-Institut),}
\centerline{\it
F\"ohringer Ring 6, D-80805 M\"unchen, Germany}

\vspace*{1.2cm}

\begin{abstract}
We construct a supersymmetric SU(5) $\times$ $A_4$ flavour GUT model in which an inverse neutrino mass hierarchy is realised without fine-tuning of parameters. 
The model shares some properties with the normal hierarchy model which we presented in~arXiv:1305.6612~-- in particular the relation $\theta_{13}^\mns \simeq  \theta_C / \sqrt{2}$. Besides these shared features, there are also important differences, mainly due to the different neutrino sector. These differences not only change the predictions in the lepton sector, but also in the quark sector, and will allow to discriminate between the two models using the results of present and future experiments. From a Markov Chain Monte Carlo fit we find that the inverse hierarchy model is in excellent agreement with the present experimental data. 
\end{abstract}

\end{titlepage}
\newpage

\section{Introduction}

Remarkable progress has been made recently regarding the measurement of the parameters in the lepton sector.
The arguably most spectacular result was the determination of a non-zero value of the reactor mixing angle $\theta_{13}^\mns$~\cite{t13_data}.
Despite these successes, the mass-ordering, the absolute mass-scale, the Dirac CP-phase and -- if neutrinos are Majorana particles -- the Majorana phases remain unmeasured to date. While the already rather precise data in the lepton sector can be regarded as a challenge for flavour model building, the fact that various parameters are still unmeasured (or will be measured more precisely in the future) means that new models which make predictions for these quantities can be tested with the data of future experiments.

It is an interesting fact that in the framework of supersymmetric (SUSY) Grand Unified Theories (GUTs) the vast majority of models features a normal neutrino mass hierarchy (NH), whereas only few such models have an inverted neutrino mass hierarchy (IH). 
Given this observation, one may ask why this is the case.
Are IH models generically more intricate and harder to construct? Or do they have more problems than NH models in fitting to the available data? 
While the above questions are hard to answer in generality, of course, we aim to answer them at least for supersymmetric flavour GUT models with an SU($5$) gauge symmetry and an $A_4$ family symmetry \cite{A4}, by constructing an example of a predictive and relatively simple IH~model of this kind and comparing it to the model with a NH but otherwise similar properties which was presented in Ref.~\cite{nh_model}.

While they have different mass hierarchies, the presented IH model shares the following features with the model of Ref.~\cite{nh_model}:
(i) the explanation of the relation $\theta_{13}^\mns \simeq  \theta_C / \sqrt{2}$ via charged lepton mixing effects linked to the Cabibbo angle by GUT relations following the strategy of \cite{Antusch:2012fb},\footnote{As a phenomenological possibility, the relation $\theta_{13}^\mns = \theta_C / \sqrt{2}$ was mentioned already some time ago~\cite{Minakata:2004xt}. Its possible origin from charged lepton corrections in Pati-Salam models has been discussed in \cite{Antusch:2011qg,King:2012vj}. In SU(5) GUTs, predictions for large $\theta_{13}^\mns$ from charged lepton corrections with consistent quark-lepton mass relations were studied in \cite{Antusch:2011qg,Marzocca:2011dh}, and conditions for realising $\theta_{13}^\mns = \theta_C / \sqrt{2}$ were given in~\cite{Antusch:2012fb}.} 
(ii) the spontaneous breaking of CP-symmetry by CP-violating vacuum expectation values (VEVs) of flavon fields leading to $\alpha \simeq 90^\circ$ in the quark unitarity triangle \cite{Antusch:2009hq,Antusch:2011sx}, and, (iii), a novel combination of Clebsch-Gordan factors from GUT symmetry breaking which leads to promising relations between quark and lepton masses.
Besides these shared features, there are also important differences, mainly due to the different neutrino sector. These differences not only change the model predictions in the lepton sector, but also affect the quark sector predictions. Interestingly, it turns out that while the presented IH model has roughly the same level of intricacy as the NH model of Ref.~\cite{nh_model}, it gives an even better fit to the data, with a $\chi^2$/d.o.f. of only $1.1$.

The paper is structured as follows: 
In section~2 we briefly review how to obtain an IH from the type~I seesaw mechanism without fine-tuning.
The superpotential for the matter sector, the required VEVs of the flavon fields and the resulting Yukawa matrices and right-handed (RH) neutrino mass matrix are given in section~3. 
In section~4 we discuss the superpotential which gives rise to the desired flavon VEVs.
Here we make use of a novel mechanism which yields $A_4$~triplet VEVs with one element vanishing and the other two elements having arbitrary magnitudes but fixed discrete phases.
The phenomenological implications and predictions of our model are presented in section~5, where we also compare these results to those of the NH model of Ref.~\cite{nh_model}. Finally, we summarize in section~6. 
In the appendix we present the messenger sector.

\section{Realising an inverted neutrino hierarchy}

If the neutrino mass ordering would be measured to be inverse, it would imply that at least two of the neutrino masses, namely $m_1$ and $m_2$, are almost degenerate (with mass splittings much smaller than the mass eigenvalues). 
By looking at the existing models in the literature (see e.g.~\cite{Albright:2009cn}), one could get the impression that, generically, such a small neutrino mass splitting would be unlikely to have its origin in a GUT flavour model. Approaches using symmetries to explain the small splitting often apply an effective ``special'' lepton number symmetry like $L_e - L_\mu -  L_\tau$ \cite{Barbieri:1998mq} or more generally some U(1) family symmetry. Although the implementation of such symmetries in GUTs is certainly possible (see e.g.\ \cite{Antusch:2005ca}), realising an inverse hierarchy often turns out to be more involved than realising a normal one. In this section, we outline our strategy\footnote{Other examples of SU(5) models with inverse hierarchy, following different approaches, have been constructed e.g.\ in~\cite{Tavartkiladze:2013hsa}.} for realising an inverse neutrino mass hierarchy in a simple way using discrete family symmetries, which can easily be embedded in SU(5) GUTs.   

\vspace{2mm}
{\bf Inverse hierarchy without fine-tuning:} One promising approach for explaining $m_1\approx m_2$ is to consider a (mostly) off-diagonal mass matrix~$M_R$ for two of the RH neutrinos, e.g. 
\be
M_R = \hat M_R \begin{pmatrix}
\varepsilon & 1 
\\
1& 0
\end{pmatrix},
\ee
so that the two RH neutrinos form a (quasi-)Dirac pair with (almost) equal moduli of the Majorana mass eigenvalues. The small entry $\varepsilon$ in the 1-1 position of $M_R$ lifts the degeneracy of the RH neutrino masses and is in turn also responsible for the small mass splitting between the light neutrino masses $m_1$ and $m_2$, as seen below. Using the seesaw formula~\cite{seesaw} 
\be
m_{\nu}  =
\frac{v_u^2}{2}(Y_\nu M_R^{-1} Y_\nu^T)\;,
\ee
where $v_u=246\;\GeV\cdot \sin\beta$, one gets with the above $M_R$ and with a neutrino Yukawa matrix (cf.~\cite{King:2000ce})
\be
Y_\nu = \begin{pmatrix}
a & 0 \\
0 & b \\
0 & c
\end{pmatrix} 
\ee
the light neutrino mass matrix
\be
\label{eq:mnu}
m_{\nu}  
=
\begin{pmatrix}
0 & B & C \\
B & 0 & 0 \\
C & 0 & 0 
\end{pmatrix}
+ \ \alpha \
\begin{pmatrix}
0 & 0 & 0 \\
0 & B & C \\
0 & C & C^2/B
\end{pmatrix},\;\text{with} \; B = b \, \frac{a \, v_u^2}{2 \hat{M}_R}, C = c \, \frac{a \, v_u^2}{2 \hat{M}_R} , \alpha = -\varepsilon \frac{b}{a}. 
\ee
The first part leads to two light neutrinos $\nu_1,\nu_2$ with exactly degenerate masses~\cite{Barbieri:1998mq}, while the second part can be considered as a small perturbation which induces the solar mass splitting. Since we consider only two RH neutrinos, the mass of the lightest neutrino is zero, $m_3 = 0$, and we have what is called a `strong inverse neutrino mass hierarchy' although strictly speaking the hierarchy is only between $m_3$ and $m_1 \approx m_2$. 

\vspace{2mm}
{\bf Implementation in flavour models:}  It is now easy to imagine how an inverse hierarchy can be realised with discrete family symmetries. For example, the columns of the neutrino Yukawa matrix can be generated by the VEVs of flavon fields which are in the $\mathbf{3}$ representation of a non-Abelian discrete family symmetry group like $A_4$.
With the flavon VEVs pointing in specific directions in flavour space, i.e.\ along $(a,0,0)^T$ and $(0,b,c)^T$, and with RH neutrinos being charged under the model's symmetry such that only the off-diagonal mass matrix entries are generated by the leading order operators, the desired structure can readily be obtained. A superpotential which makes the flavon VEVs point in the above described directions will be presented in section~\ref{sec:alignment}. 

\vspace{2mm}
{\bf Mixing angles and the Dirac CP phase: }
It is straightforward to calculate the two non-vanishing mass eigenvalues $m_i$ and the mixing angles $\theta^{\nu}_{ij}$ which bring $m_{\nu}$ into diagonal form:\footnote{Three comments are in order: (i) The subscripts of the eigenvalues correspond to the standard mass ordering for $\alpha > 0$. (ii) We assume all parameters to be real. (iii) The symbol `$\approx$' here means that the result is in leading order in $\alpha$.} 
\besub
\bal
-\Delta m^2_{\text{atm}} = m_2^2 &\approx B^2+C^2 \,,\\
\Delta m^2_{\text{sol}} = m_2^2-m_1^2&  \approx 2 \, \alpha \, \frac{(B^2+C^2)^{3/2}}{|B|}\,, \\
\tan \theta_{12}^{\nu}&\approx \left\vert 1 - \frac{\alpha}{2} \, \frac{\sqrt{B^2+C^2}}{|B|} \right\vert \approx
    \left\vert 1 + \frac{1}{4} \frac{\Delta m^2_{\text{sol}}}{\Delta m^2_{\text{atm}}}\right\vert\,, \\
\tan \theta_{23}^{\nu}&=\left\vert C/B \right\vert \,,\\
\theta_{13}^{\nu}&=0 \,.
\eal
\eesub
As can be seen, the relative smallness of the mass splitting between $\nu_1$ and $\nu_2$ directly results from the smallness of the correction $\alpha$ (which stems from a higher-dimensional operator) and is not due to a cancellation of a priori unrelated parameters.

Additionally, the mixing angle $\theta_{23}^{\nuL}$ is a free parameter.
On the other hand, the deviation from $\tan \theta_{12}^{\nuL}=1$ is smaller than $10^{-2}$, which amounts to a negligibly small deviation from $\theta_{12}^{\nuL} = 45^\circ$, and the 1-3 mixing in the neutrino sector is fixed to zero. 

The fact that $\theta_{13}^{\nuL} = 0$ opens up the interesting possibility of explaining the PMNS angle $\theta_{13}^\mns$ via charged lepton mixing effects in GUTs, following the strategy described in \cite{Antusch:2012fb}. When the charged lepton Yukawa matrix $Y_e$ is diagonalised mainly by a 1-2 rotation $\theta_{12}^{e}$ connected by GUT relations to $\theta_{12}^{e} \simeq  \theta_{12}^{d} \simeq \theta_C$ (in the rest of this section we will assume $\theta_{13}^{e} \approx 0$ and $\theta_{23}^{e}  \approx 0$), then $\theta_{13}^\mns$ is generated with a value in very good agreement with experiments:
\be\label{eq:t13fromt12e}
\theta_{13}^\mns \simeq \theta_{12}^{e} \sin (\theta_{23}^\mns)   \simeq  \theta_C / \sqrt{2}  \:.
\ee

In addition, such a charged lepton mixing contribution also affects the PMNS angle $\theta_{12}^\mns$, as can be seen from the lepton mixing sum rule~\cite{sumrule}
\be
    \theta_{12}^\mns  \simeq \theta_{12}^\nu +\theta_{12}^{e} \cos(\theta_{23}^\mns)\cos(\delta^\mns)\;. 
\label{eq:sumrule}
\ee
One can also see that $\theta_{12}^\mns$ can be in very good agreement with the experimental data as long as $\delta^{\mns} \simeq 180^\circ$, and as long as the angle $\theta_{23}^\mns$ is smaller than $45^\circ$, more precisely $\theta_{23}^\mns \approx 40^\circ$, so that $\cos(\theta_{23}^\mns)$ is somewhat larger than $1/\sqrt{2}$.\footnote{Note that with $\theta_{23}^\mns = 45^\circ$, $\theta_{12}^\mns$ would be about $36^\circ$, which is disfavoured by more than $3\sigma$.} 
Note that $\theta^e_{13}\approx\theta^e_{23}\approx0$, $\delta^\mns=180^\circ$ and $\theta_{23}^\mns<45^\circ$ will be built-in features of the model, as we discuss in the following section.

\section{The model}
\label{sec:model}

We now turn to the actual construction of the model. The first step towards this end is the identification of effective operators that lead to the desired Yukawa matrices after the GUT symmetry is broken and after the flavons attain their VEVs.
This will be treated in this section.
In section~4 we will discuss the superpotential that gives rise to the flavon VEVs.
Note that, to make the model consistent, one has to identify its `shaping symmetry' and furthermore find a set of messenger fields, that, when integrated out, give rise to the effective operators in the superpotential.
It is crucial that the shaping symmetries and messengers do not allow any further effective operators which could spoil the features of our Yukawa matrices, e.g.~by generating unwanted Clebsch factors or relevant corrections to the desired structure of the Yukawa matrices.

The matter content of the Minimal Supersymmetric Standard Model (MSSM) is contained (using the standard SU(5) embedding) in the $\mathbf{\bar 5}$~-- which in our model is an $A_4$ triplet~-- and three $\bf 10$'s of SU(5) that are $A_4$-invariant singlets.
As discussed in the previous section, we introduce two right-handed neutrinos $N_1$ and $N_2$ which are SU(5) and $A_4$ singlets. 
To obtain the Clebsch-Gordan factors $6$, $-1/2$ and $-3/2$~-- which are useful in order to get a model with a good fit to the data, as was already discussed in Ref.~\cite{nh_model}~-- we introduce a Higgs fields $H_{\overline{45}}$ in the $\mathbf{\overline{45}}$ of SU(5), which, together with the usual $H_5$ and $H_{\bar{5}}$, is responsible for the breaking of the electroweak symmetry. 
The GUT gauge group gets broken to the SM gauge group by the VEV of a Higgs field $H_{24}$ in the $\mathbf{24}$ of SU(5).\footnote{While we explicitly discuss the matter sector and the breaking of the family symmetry, the details of the Higgs potential are beyond the scope of this paper.}
When SU(5) gets broken Clebsch-Gordan ratios between the charged leptons and down-type quarks arise, depending on the contraction between $H_{24}$ and the matter fields. In order to ensure that only the desired Clebsch factors appear in the model, one needs to explicitly construct the renormalizable theory with messenger fields. In Table~\ref{tab:A4messengerSector} on pages \pageref{tab:A4messengerSector}ff we present the messenger sector and show the operators that lead to the effective superpotential after the messenger fields are integrated out.
 
The rows respectively columns of the Yukawa matrices are formed by specific VEVs of flavon fields. Five flavons of the model are $A_4$-triplets and three are $A_4$-invariant singlets. 
In Table~\ref{tab:flavonvevs} we present their VEVs. 
Note that all parameters $\epsilon_i$ are real, as explained in the next section.
For convenience we introduce a single symbol $\Lambda$ for the suppression by messenger mass scales, keeping in mind that in fact each effective operator is in general suppressed by different messenger masses, as can be seen from the figures on pages \pageref{tab:A4messengerSector}ff.
\begin{table}[!h]
\centering 
\begin{tabular}{|c|| c| c| c| c| c| c| c| c|}
\hline
\rule{0 px}{15 px}
flavon $\varphi_i$:
&$\phi_1$
&$\phi_2$
&$\phi_3$
&$\phi_{ab}$
&$\phi_{bc}$
&$\xi_{12}$
&$\xi_{23}$
&$\xi_{M}$ \\
\hline
\rule{0 px}{30 px}
$\frac{\vev{\varphi_i}}{\Lambda}$:
&$\epsilon_1 \begin{pmatrix} 1 \\ 0 \\ 0 \end{pmatrix}$
&$\epsilon_2 \begin{pmatrix} 0 \\ -i \\ 0 \end{pmatrix}$
&$\epsilon_3 \begin{pmatrix} 0 \\ 0 \\ 1 \end{pmatrix} $
&$\epsilon_{ab} \begin{pmatrix} c_{ab} \\ s_{ab} \\ 0 \end{pmatrix}$
&$\epsilon_{bc} \begin{pmatrix} 0\\ c_{bc} \\ s_{bc} \end{pmatrix}$
&$\epsilon_{12}$
&$\epsilon_{23} $&$\epsilon_{M}$\\[4ex]
\hline
\end{tabular}
\caption{The VEVs of the flavon fields. The $\epsilon_i$ are all assumed to be real numbers. See the following section for a discussion of the flavon alignment. We abbreviated $c_{ab}\equiv \cos (\theta_{ab})$ and $s_{ab}\equiv \sin (\theta_{ab})$ and analogously for $\theta_{bc}$.} \label{tab:flavonvevs}
\end{table}

The effective superpotential for the matter sector is given by (up to real coefficients and omitting messenger mass scales $\Lambda$ for the sake of brevity) $W=W_{Y_{\nu}} +W_{M_R} +W_{Y_d} +W_{Y_u}$ with\footnote{We denote the decomposition into the SU(5) representation $\mathbf{R}$ by $[\dots]_{R}$.}
\besub
\label{eq:superpotential}
\begin{align}
W_{Y_{\nu}} &=  (H_5 F) ( N_1\phi_1 + N_2\phi_{bc} ) \label{eq:wnu}\,,\\
W_{M_R} &= \xi_M^4 ( N_1 N_2  + \phi_{bc}^2 N_1^2  ) \,, \\
W_{Y_d} &= [T_1  H_{\overline{45}}]_{45} [F H_{24}]_{\overline{45}} \ \phi_2
+ [T_2 H_{24}]_{10} [F H_{\bar 5}]_{\overline{10}}  \ \phi_{ab}  
+ [T_3 H_{\bar 5}]_5 [F H_{24}]_{\bar{5}} \ \phi_3 \,,\\
W_{Y_u} &= H_5(T_3^2  +  T_2^2 \phi_{ab}^2+  T_1^2 (\phi_2^2)^2  + T_2T_3 \xi_{23} + T_1 T_2 \xi_{12}^5  ) \,.
\end{align}
\eesub
After GUT- and family symmetry breaking the Yukawa matrices of the quarks and charged leptons are given by
\begin{equation} \label{YdYeYu}
    Y_d = 
    \begin{pmatrix}
        0 & i\tilde \epsilon_2 & 0 \\
        \tilde \epsilon_{ab} c_{ab} &  \tilde \epsilon_{ab} s_{ab} & 0\\
        0 & 0 & \tilde \epsilon_3
    \end{pmatrix},\;
    Y_e = 
    \begin{pmatrix}
        0 & 6 \tilde \epsilon_{ab} c_{ab} & 0 \\
        - i\frac 1 2 \tilde \epsilon_2 & 6 \tilde \epsilon_{ab} s_{ab} & 0	 \\
        0 & 0& -\frac{3}{2} \tilde \epsilon_3
    \end{pmatrix},\;
    Y_u = \begin{pmatrix}
        \epsilon_2^4 & \epsilon_{12}^5 & 0 \\
        \epsilon_{12}^5 &\epsilon_{ab}^2 & \epsilon_{23} \\
        0 &  \epsilon_{23} &  y_t
    \end{pmatrix},
\end{equation}
where we defined
\be
\tilde\epsilon_i = \frac{\langle H_{24}\rangle}{\Lambda}\epsilon_i
\ee
for convenience. Sub-leading corrections, e.g. from higher-dimensional operators or from canonical normalisation~\cite{CN} can be neglected with the messenger sector of the model presented in the Appendix. Note that we use the convention of the Particle Data Group~\cite{pdg}
\be
W = (Y_u^*)_{ij} Q_i u^c_j H_u + (Y^*_d)_{ij} Q_i d^c_j H_d + (Y^*_e)_{ij} L_i e^c_j H_d  + (Y^*_\nu)_{ij} L_i \nu^c_j H_u  + (M_{\nu^c}^*)_{ij} \nu^c_i \nu^c_j\;,
\ee
so that the complex conjugates of the flavon fields form the rows and columns of the down-type quark and charged lepton Yukawa matrices, respectively. 

Finally, the desired neutrino Yukawa matrix and right-handed neutrino mass matrix~-- which, as discussed in section~2, lead in a natural way to an inverse neutrino mass hierarchy~-- are obtained from $W_{Y_\nu}$ and $W_{M_R}$. The parameters $\hat M_R$ and $\varepsilon$ in Eq.~(\ref{eq:mnu}) are then given in terms of $\epsilon_{bc}$ and $\epsilon_M$.

\section{Flavon alignment}\label{sec:alignment}

Having discussed the matter sector of the model we now present the superpotential responsible for the vacuum alignment of the $A_4$-triplet flavons $\phi$ and $A_4$-singlet flavons $\xi$.
The scalar potential for the flavons is obtained from the $F$-terms of the `driving fields' $S$, $D$, $A$, $O$ and $O^\prime$. 
All flavons and driving fields are listed in the lower part of Table~\ref{tab:A4MatterflavSector} together with their charges under the imposed symmetries. 

The total effective superpotential which is responsible for the vacuum alignment of the flavons is given by
\be \label{Wflavon1}
W_{\textrm{flavon}}=W_{\phi_i} + W_{ab,23} + W_{bc}+ W_M + W_{12}
\ee
where\footnote{For products of $A_4$ representations, we use the Ma-Rajasekaran (``$SO(3)$-like'') basis, as introduced in the first reference of~\cite{A4}. The product $\mathbf{3}\otimes \mathbf{3}$ decomposes into $\mathbf{1}\oplus \mathbf{1^\prime} \oplus \mathbf{1^{\prime\prime}}\oplus \mathbf{3_s} \oplus \mathbf{3_a}$. The invariant singlet $\mathbf{1}$ is given by the $SO(3)$-type inner product. $\mathbf{3_s}$ is constructed from the symmetric star-product ($\star$) and $\mathbf{3_a}$ from the antisymmetric cross-product ($\times$), respectively. We denote the contraction to the $\mathbf{1^\prime}$ and $\mathbf{1^{\prime\prime}}$ representations with $(\dots)_{1^\prime}$ and $(\dots)_{1^{\prime\prime}}$, respectively.}
\besub \label{Wflavon2}
\bal
W_{\phi_i}  &= 
S_2[(\phi_2^2)^3+M_2^2]+\sum_{i\in\{1,3\}}S_i[\phi_i^2-M_i^2]+\sum^3_{\substack{i,j=1\\j>i}}\left[O^\prime_{i;j}(\phi_i\phi_j)_{1^{\prime\prime}}+O_{i;j}(\phi_i\phi_j)\right] \,,  \\
W_{ab,23} 
&= S_{ab} [ (\phi_{ab}\star\phi_{ab})^2 \xi_{23}^2 -M_{ab23}^2 ] 
+D_{ab}^{\alpha}[ \phi_{ab}^2+ \lambda_{23 }\xi_{23}^2]  \nn \\
& \qquad + D_{ab}^{\beta} (\phi_{ab}\star\phi_{ab})\phi_{ab}
+ D_{ab}^{\gamma} [ (\phi_{ab}^2)_{1'} (\phi_{ab}^2)_{1''}+ k_{ab} \ (\phi_{ab}\star\phi_{ab})^2 ]  \,,  \\
W_{bc}
&= S_{bc} [ (\phi_{bc}\star\phi_{bc})^2 \phi_{bc}^2-M_{bc}^2 ]
+ D_{bc}^{\beta} (\phi_{bc}\star\phi_{bc})\phi_{bc}
+ D_{bc}^{\gamma} [(\phi_{bc}^2)_{1'} (\phi_{bc}^2)_{1''}+ k_{bc} (\phi_{bc}\star\phi_{bc})^2 ] \,, \\
W_{M} &=
S_{M}[\xi_{M}^6-M_{M}^2] \,,  \\
W_{12} &=
S_{12}[\xi_{12}^6-M_{12}^2] \,.
\eal
\label{eq:align}
\eesub
Since CP symmetry is assumed to be broken only spontaneously, the constants $k_i$ and $M_i$ are real. In addition we assume $M_i^2>0$ everywhere. The terms $W_{12}$ and $W_M$ lead to VEVs with discrete phases as discussed in \cite{Antusch:2011sx}, of which we select the real VEVs.

Let us elaborate on the flavon alignment $W_{\phi_i}$. The $F$-term equations corresponding to $O_{i;j}$ and $O^\prime_{i;j}$ take the form
\besub
\begin{align}
    x_1 + x_2 + x_3 &= 0 \;,\\
    x_1 + x_2 \,\omega^2 + x_3 \, \omega &= 0 \;,
\end{align}
\eesub
where $\omega = \exp(i \, 2\pi/3 )$ and $x_k = (\phi_i)_k (\phi_j)_k$.
There is one such system of equations for each pair $O_{i;j}$ and $O^\prime_{i;j}$. Their solutions  can be grouped into two classes: (i) $x_1 = x_2 = x_3 = 0$ and (ii) $x_1 = \gamma, x_2 = \omega^2 \, \gamma, x_3 = \omega \, \gamma$, with a complex constant $\gamma \not= 0$. When translating these two classes into solutions for the components $(\phi_i)_k$, one finds that no combinations mixing (i) and (ii) are consistent. Thus the only solutions to all three equation pairs are either all of type (i) or of type (ii). They read 
\begin{enumerate}
    \item[(i)] $\phi_1 \propto (1,0,0)^T, \phi_2 \propto (0,1,0)^T, \phi_3 \propto (0,0,1)^T$ or permutations thereof, 
    \item[(ii)] $\phi_i \propto (1,\omega,\sqrt{\omega})^T$ for $i = 1,2,3$.
\end{enumerate}
Which solution is realised depends on the type of $F$-term that is used to force $\phi_i \ne 0$. Namely, for solution (ii) the invariant $\phi_i^2$ vanishes for all $\phi_i$, so that in our setup only solution (i) is selected. Again their phases take on discrete values. We select an imaginary VEV for $\phi_2$ and real VEVs for $\phi_1$ and $\phi_3$, as  presented in Table~\ref{tab:flavonvevs}.

Finally, we briefly consider $W_{ab23}$ and $W_{bc}$. In general, the first term (with the singlet driving fields $S_{ab}$, $S_{bc}$) determines the magnitude\footnote{Note that the $F$-term of $S_{ab}$ does not fix the individual magnitudes of $\phi_{ab}$ and $\xi_{23}$, but only the magnitude of their product. The individual magnitudes are determined by the $F$-term of $D_{ab}^{\alpha}$.} of the VEV and the global phase is set to discrete values due to the real value of $M^2$. The second term (with the driving field $D_{ab}^{\beta}$, $D_{bc}^{\beta}$) sets one of the components of $\langle\phi \rangle$ to zero. For the two remaining entries of the flavon, the relative magnitude or the relative phase is set by the value of $k_{ab}$, $k_{bc}$. Let us exemplify this for $\langle \phi_{bc} \rangle$ ($\langle \phi_{ab} \rangle$ is determined in a completely analogous way). Because of the terms with $S_{bc}$ and $D_{bc}^\beta$ the global phase and magnitude of $\langle \phi_{bc} \rangle$ is fixed and one component is known to vanish. We therefore write
\be\label{eq:phibc}
\langle \phi_{bc} \rangle = \begin{pmatrix} 0 \\ b \\ c \end{pmatrix} \,.
\ee
Setting the $F$-term of $D_{bc}^\gamma$ to zero yields three different types of solution depending on $k_{bc}$. For $-1 \leq k_{bc} \leq 3$, one has $|c/b|=1$ and the relative phase $\varphi = \text{arg}(c / b)$ can be fixed by $k_{bc} = 1 - 2 \cos(2 \varphi)$. For $k_{bc} < -1$ and $k_{bc} > 3$, the situation is reversed and $k_{bc}$ governs the ratio $c/b$ and the phase $\varphi$ is fixed to $\varphi \in \{0, \pi\}$ for $k_{bc} < -1$ and $\varphi = \pm 90^\circ$ for $k_{bc} > 3$. 
Defining $\tan \theta = |c/b|$ with $\theta \in [0^\circ,90^\circ)$, one can obtain a specific value of $\theta$ by choosing $k_{bc} = -1 - 4 \cot^2(2 \theta)$ for $k_{bc} < -1$ or $k_{bc} = 3 + 4 \cot^2(2 \theta)$ for $k_{bc} > 3$. The same formulae also apply to $\phi_{ab}$ with $b,c \to a,b$, with $a$ and $b$ defined analogously to Eq.~\eqref{eq:phibc}. 
For both $\phi_{bc}$ and $\phi_{ab}$ we select the real VEVs.
 
Note that with the flavon alignment of Eq.~(\ref{eq:align}) all driving- and flavon fields get masses of the order of the family symmetry breaking scale.

\section{Phenomenology}

In this section we present a numerical fit of the model to the measured observables, a Markov Chain Monte Carlo (MCMC) analysis and a discussion of the model's predictions. We perform the numerical analysis in a similar way as in Ref.~\cite{nh_model}. Because of the differences in the neutrino sector and for the sake of completeness, we nevertheless briefly describe the numerical procedure.

Our model has 14 free parameters: $\tilde \epsilon_2$, $\tilde \epsilon_3$, $\tilde \epsilon_{ab}$, $\theta_{ab}$, $\eta_{Q_{12}}$ and $\eta_{Q_{3}}$ parametrize the Yukawa matrices $Y_d$ and $Y_e$. The parameters $y_t$, $\epsilon_{ab}$, $\epsilon_2$, $\epsilon_{12}$ and $\epsilon_{23}$ enter $Y_u$, where we fix $\epsilon_{ab}$ and $\epsilon_2$ to be positive, which is possible since they appear at quadratic and quartic order, respectively. Since there are only three relevant parameters in $m_\nu$, as discussed in section 2, we can keep two parameters of the neutrino sector fixed. While  $Y_\nu$ is parametrized by $\epsilon_1$, $\epsilon_{bc}$ and $\theta_{bc}$, we set the parameters of $M_R$ to $\hat M_{R}=10^{10}\ \GeV$ and $\varepsilon=10^{-2}$. Note that for neutrino Yukawa couplings much smaller than 1, which implies that $Y_\nu$ is irrelevant for the renormalization group (RG) evolution, the choice of $M_R$ does not affect the fit. The neutrino sector is therefore given by
\be
Y_\nu = \begin{pmatrix}
\epsilon_{1} &  0 \\
0 &  \epsilon_{bc}\cos{\theta_{bc}}\\
0  &  \epsilon_{bc}\sin{\theta_{bc}}
\end{pmatrix}\;,
\quad
M_R = 
10^{10}~\text{GeV}~\cdot
\begin{pmatrix} 
 10^{-2} & 1 \\
1 &   0
\end{pmatrix}\,.
\label{Yukawa-pheno}
\ee
With this parametrization we perform a fit of the 14 parameters to the 18 measured observables. In addition to the 9 fermion masses, 6 CKM and PMNS mixing angles, the CKM phase and the  two neutrino mass splittings, we also calculate the yet unmeasured Dirac CP phase $\delta^\mns$ and the Majorana phase $\varphi^\mns=\varphi_2^\mns-\varphi_1^\mns$, which is the single physical Majorana phase for the case of a massless lightest neutrino $m_3$ (we use the notation of the Mathematica package REAP~\cite{Antusch:2005gp}). Our model therefore makes 6 predictions.   

We use the one-loop MSSM RGEs and REAP to run the Yukawa matrices from the GUT scale $M_{\mathrm{GUT}}= 2\cdot 10^{16}~\GeV$ to the superpartner mass scale $\Lambda_\mathrm{SUSY}=1\ \TeV$. 
At their respective mass scales, the RH neutrinos are integrated out and the effective mass matrix of the light neutrinos is calculated. In order for the Clebsch-Gordan ratios $6$ and $-\tfrac{3}{2}$ to be viable, a large or moderate $\tan\beta$ is required~\cite{Antusch:2009gu}. 
We set $\tan\beta=40$ and include the $\tan\beta$-enhanced SUSY threshold corrections~\cite{SUSYthresholds,Blazek} 
at the matching scale of the MSSM to the Standard Model (SM), which are given by the approximate relations
\besub\label{eq:SUSYthreshold}
\begin{align}
Y_d^{\text{SM}} &= (\mathbf{1} + \text{diag} (\eta_{Q_{12}},\eta_{Q_{12}},\eta_{Q_{3}})) \, Y_d^{\text{MSSM}} \cos\beta\;,\\
Y_{u}^{\text{SM}} &= Y_{u}^{\text{MSSM}}\sin\beta\;,\\
Y_e^{\text{SM}} &= Y_e^{\text{MSSM}}\cos\beta\;,
\end{align}
\eesub
in the basis where $Y_u$ is diagonal. The $\eta_i$ are proportional to $\tan\beta$ and treated as free parameters in the fit. In an explicit SUSY scenario they could be calculated from the sparticle spectrum and $\tan\beta$. We restrict the absolute values $|\eta_i|<0.5$, which is justified by the size of the SUSY threshold corrections in common SUSY breaking scenarios (see e.g.~\cite{Antusch:2008tf}). We remark that we have absorbed the SUSY threshold corrections for the charged leptons in the quark corrections. This is valid to a good approximation, since GUTs only predict ratios of quark and lepton masses.

Below $\Lambda_\mathrm{SUSY}$ we use the Standard Model one-loop RGEs to run the Yukawa matrices to the mass scale of the top quark $m_t(m_t)=162.9\; \GeV$ where all observables are calculated and compared with results from measurements. The values of the charged lepton- and quark masses are taken from Ref.~\cite{Xing:2007fb}. Note that because of the accuracy of roughly one percent of the one-loop calculations used in the fit, we set the uncertainty of the charged lepton masses to one percent, although they are given to higher precision in Ref.~\cite{Xing:2007fb}. We use the Winter 2013 fit results of the UTfit collaboration~\cite{UTfit} for experimental data of the CKM parameters and the updated global fit results of the NuFIT collaboration~\cite{GonzalezGarcia:2012sz} for the lepton mixing observables.

\subsection{Results}

Using this procedure, we find a best fit point with a total $\chi^2=4.6$. The parameter values corresponding to this are presented in Table~\ref{tab:pararesults}. Having 14 parameters and 18 fitted observables we have a reduced $\chi^2$ of $\chi^2/\mathrm{d.o.f.}=1.1$. Compared to $\chi^2/\mathrm{d.o.f.}=2.0$ in the fit of the normal hierarchy model of Ref.~\cite{nh_model}, we obtain an even better agreement with the measured observables.
Additionally, we perform a MCMC analysis with a Metropolis-Hastings algorithm and give $1\sigma$ highest posterior density (HPD) intervals~\cite{pdg} as an uncertainty. 
\begin{table}[!h]
\centering 
\begin{tabular}{|rl|r@{.}l|c|}
\hline
\multicolumn{2}{|c|}{Parameter} & \multicolumn{2}{|c|}{Best fit value} & Uncertainty\\
\hline
\hline 
\rule{0 px}{15 px}
$\tilde \epsilon_2$ & in $10^{-4}$ &\hspace{.55cm} $6$&$71$ & $_{-0.07}^{+0.08} $ \\
\rule{0 px}{15 px}
$\tilde \epsilon_3$ & in $10^{-1}$& $2$&$23$ & $\pm 0.03 $\\
\rule{0 px}{15 px}
$\tilde \epsilon_{ab}$ & in $10^{-3}$ & $ 3$&$03$ & $\pm 0.03 $ \\
\rule{0 px}{15 px}
$\theta_{ab}$ & & $ 1$&$314$ & $_{-0.003}^{+0.004} $\\
\rule{0 px}{15 px}
$\eta_{Q_{12}}$ & in $10^{-1}$ & $ 3$&$36$ & $_{-2.50}^{+1.35} $ \\
\rule{0 px}{15 px}
$\eta_{Q_3}$ & in $10^{-1}$ & $ 1$&$64$ & $_{-0.37}^{+0.44} $ \\ 
\rule{0 px}{15 px}
\multirow{2}{*}{$\epsilon_2$} & 
\multirow{2}{*}{in $10^{-2}$} & 
\multicolumn{2}{|c|}{\multirow{2}{*}{$\begin{cases}5.22\\[2px]6.04 \end{cases}$}} & 
\multirow{2}{*}{$\begin{cases}_{-0.37}^{+0.34} \\[2px]_{-0.23}^{+0.18} \end{cases}$} \\ 
\rule{0 px}{15 px}
& & \multicolumn{2}{|c|}{} &\\
\rule{0 px}{15 px}
$\epsilon_{ab}$ & in $10^{-2}$ &  $ 4$&$45$ & $_{-0.18}^{+0.53} $\\
\rule{0 px}{15 px}
$\epsilon_{12}$ & in $10^{-1}$  & $ -1$&$64$ & $_{-0.04}^{+0.05} $\\
\rule{0 px}{15 px}
$\epsilon_{23}$ & in $10^{-2}$ &  $ 1$&$76$ & $_{-0.16}^{+0.49} $\\
\rule{0 px}{15 px}
$y_t$ & in $10^{-1}$ & $ 5$&$30$ & $_{-0.27}^{+0.35} $\\
\rule{0 px}{15 px}
$\epsilon_1$ & in $10^{-3}$   & $ 3$&$21$ & $_{-0.09}^{+0.11} $\\
\rule{0 px}{15 px}
$\epsilon_{bc}$ & in $10^{-3}$ & $ -5$&$12$ & $^{+0.11}_{-0.15} $ \\
\rule{0 px}{15 px}
$\theta_{bc}$ &   & $ 0$&$71$ & $\pm 0.02 $\\[2px]
\hline
\end{tabular}
\caption{Best fit results of the parameters with $\chi^2/\mathrm{d.o.f.}=1.1$. The uncertainties are given as $1\sigma$ highest posterior density intervals. The two modes for $\epsilon_2$ can be understood as the two solutions of the (leading order) equation  $y_u\approx\vert(Y_u)_{11}-(Y_u)_{12}^2/(Y_u)_{22}\vert$, where $(Y_u)_{11}=\epsilon_2^4$.} \label{tab:pararesults}
\end{table}

\begin{figure}[h!]
\centering
        \begin{subfigure}[b]{0.4\textwidth}
                \centering
                \includegraphics[width=\textwidth]{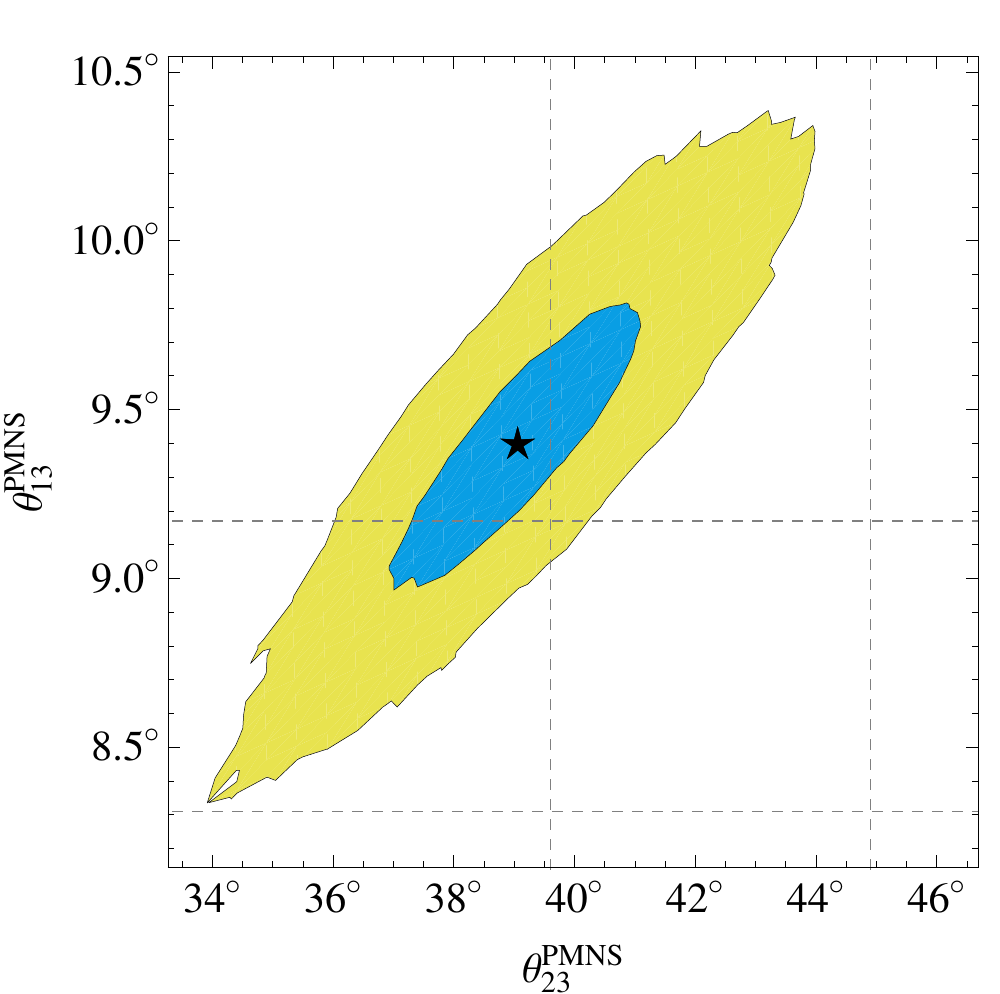}
                \vspace{-.5cm}
                \caption{}
                \label{subfig:t23t13}
        \end{subfigure}
        \quad
        \begin{subfigure}[b]{0.4\textwidth}
                \centering
                \includegraphics[width=\textwidth]{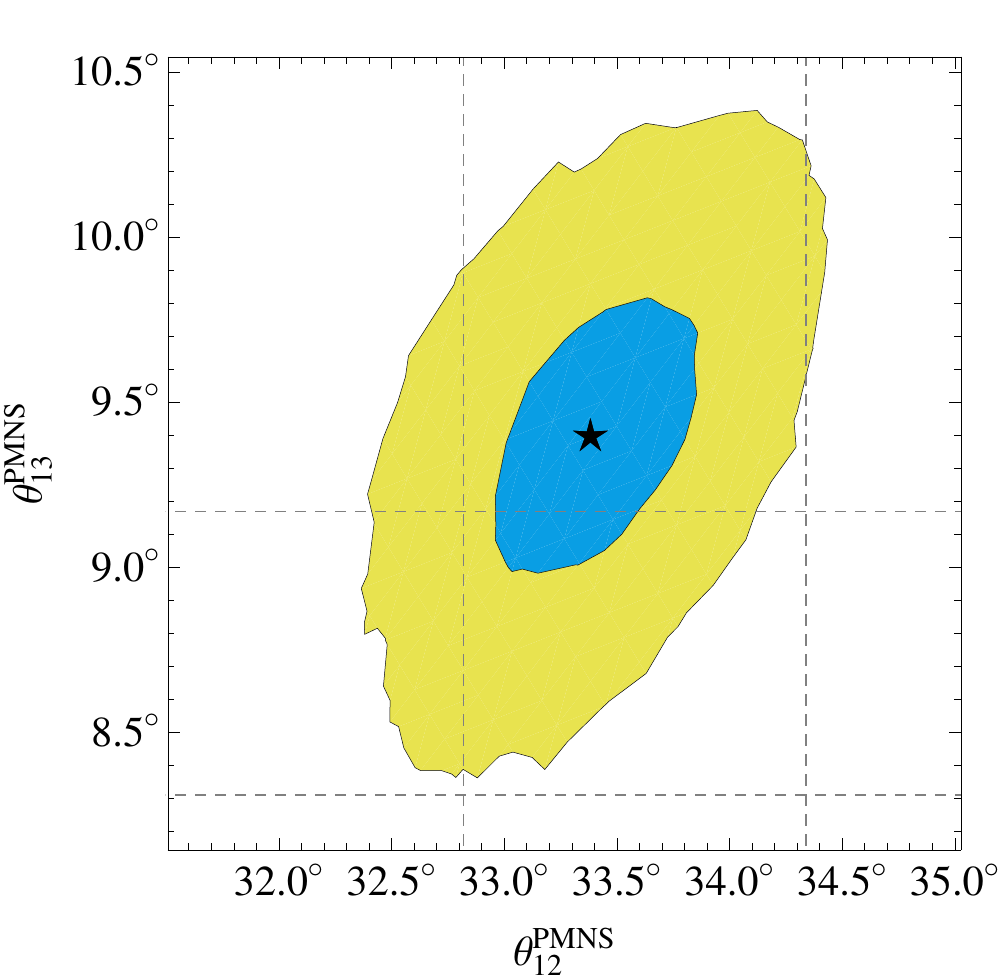}
                \vspace{-.5cm}
                \caption{}
                \label{subfig:t12t13}
        \end{subfigure}
        
        \begin{subfigure}[b]{0.4\textwidth}
                \centering
                \includegraphics[width=\textwidth]{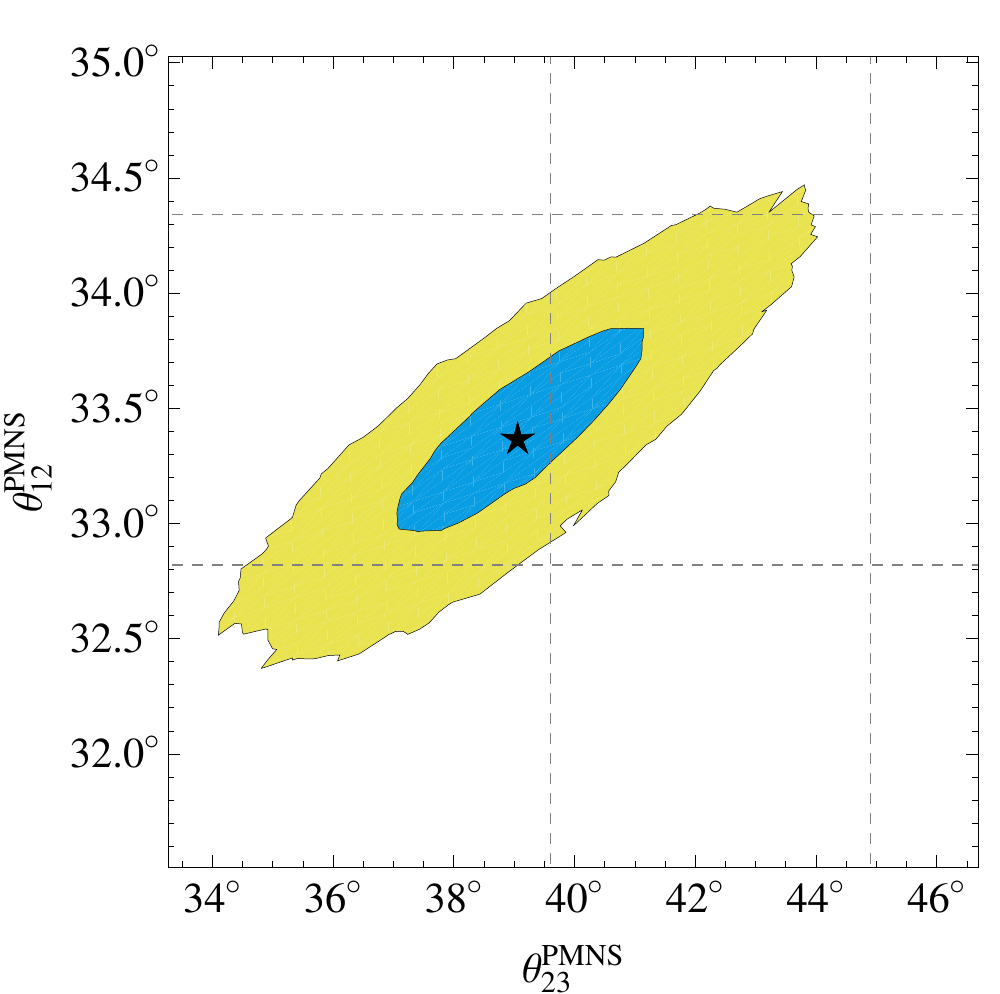}
                \vspace{-.5cm}
                \caption{}
                \label{subfig:t23t12}
        \end{subfigure}
\caption{Correlations among the lepton mixing angles. The black star marks the best fit value. The blue and golden regions give the $1\sigma$ and $3\sigma$ HPD regions obtained from the MCMC analysis, respectively. The dashed grey lines indicate the $1\sigma$ intervals of the measured observables.}
\label{fig:results}
\end{figure}

In Table~\ref{tab:results} we show the best fit values of the observables at $m_t(m_t)$. For convenience we also list the lepton mixing angles in degree in Table~\ref{tab:results2}. In Figure~\ref{fig:results} the correlations among the lepton mixing angles in the MCMC analysis are shown.

\begin{table}[t]
\centering 
\begin{tabular}{|rl|rl|r@{.}l|c|}
\hline
\multicolumn{2}{|c|}{Observable} &  \multicolumn{2}{|c|}{Value at $m_{t}$} & \multicolumn{2}{|c|}{Best fit result} & Uncertainty\\
\hline
\hline
\rule{0 px}{15 px}
$m_u$ & in MeV & $1.22$&$_{-0.40}^{+0.48}$ & $1$&$22$ &  $ _{-0.39}^{+0.50}$ \\ 
\rule{0 px}{15 px}
$m_c$ & in GeV & $0.59$&$ \pm 0.08$ & $0$&$59$ &  $ _{-0.09}^{+0.07}$\\
\rule{0 px}{15 px}
$m_t$ & in GeV & $162.9$&$ \pm 2.8$ & $162$&$91$ &  $ _{-2.44}^{+3.35}$\\[2px]
\hline
\rule{0 px}{15 px}
$m_d$ & in MeV & $2.76$&$_{-1.14}^{+1.19}$ & $2$&$73$ &  $ _{-0.54}^{+0.25}$\\
\rule{0 px}{15 px}
$m_s$ & in MeV & $52$&$ \pm 15$ & $50$&$70$ &  $ _{-9.72}^{+4.86}$\\
\rule{0 px}{15 px}
$m_b$ & in GeV & $2.75$&$ \pm 0.09$ & $2$&$75$ &  $ \pm 0.09$\\[2px]
\hline
\rule{0 px}{15 px}
$m_e$ & in MeV & $0.485$&$ \pm 1\%$ & $0$&$483$ &  $ \pm 0.005$\\
\rule{0 px}{15 px}
$m_\mu$ & in MeV & $102.46$&$ \pm 1\%$ & $102$&$87$ &  $ _{-0.91}^{+1.04}$\\
\rule{0 px}{15 px}
$m_\tau$ & in MeV & $1742$&$ \pm 1\%$ & $1741$&$99$ &  $ _{-17.70}^{+16.84}$\\[2px]
\hline
\rule{0 px}{15 px}
$\sin\theta_C$ & & $0.2254$&$ \pm 0.0007$ & $0$&$2255$ &  $ \pm 0.0007$\\
\rule{0 px}{15 px}
$\sin\theta_{23}^\ckm$ & & $0.0421$&$ \pm 0.0006$ & $0$&$0421$ &  $ \pm 0.0006$\\
\rule{0 px}{15 px}
$\sin\theta^\ckm_{13}$ && $0.0036$&$ \pm 0.0001$ & $0$&$0036$ &  $ \pm 0.0001$\\
\rule{0 px}{15 px}
$\delta^\ckm$ & in $^\circ$& $69.2$&$ \pm 3.1$ & $69$&$27$ &  $ _{-0.69}^{+0.91}$\\[2px]
\hline
\rule{0 px}{15 px}
$\sin^2\theta^\mns_{12}$ & & $0.306$&$ \pm 0.012$ & $0$&$303$ &  $ \pm 0.005$\\
\rule{0 px}{15 px}
$\sin^2\theta^\mns_{23}$ & & $0.437$&$_{-0.031}^{+0.061}$ & $0$&$397$ &  $ _{-0.022}^{+0.023}$\\
\rule{0 px}{15 px}
$\sin^2\theta^\mns_{13}$ && $0.0231$&$ _{-0.0022}^{+0.0023}$ & $0$&$0267$ &  $ _{-0.0015}^{+0.0016}$\\
\rule{0 px}{15 px}
$\delta^\mns$ & in $^\circ$& \multicolumn{2}{|c|}{-} & \multicolumn{2}{|c|}{$180$} &   -\\
\rule{0 px}{15 px}
$\varphi^\mns$ & in $^\circ$& \multicolumn{2}{|c|}{-} & \multicolumn{2}{|c|}{$180$} &  -\\[2px]
\hline
\rule{0 px}{15 px}
$\Delta m^2_{\text{sol}}$ & in $10^{-5}$ eV$^2$ & $7.45$&$ _{-0.16}^{+0.19}$ & $7$&$45$ &  $ _{-0.17}^{+0.18}$\\
\rule{0 px}{15 px}
$\Delta m^2_{\text{atm}}$ & in $10^{-3}$ eV$^2$& $-2.410$&$_{-0.063}^{+0.062}$ & $-2$&$410$ &  $ _{-0.064}^{+0.062}$\\[2px]
\hline
\end{tabular}
\caption{Best fit results and uncertainties of the observables at $m_t(m_t)$. We give $1\sigma$ highest posterior density intervals as uncertainty. $\delta^\mns$ and $\varphi^\mns$ are exactly $180^\circ$, since $Y_\nu$ is real and the phase of $(Y_e)_{21}$ can be absorbed by the right-handed electron field. Note that although the masses of the charged leptons are known far more precise than listed here, we set an $1\%$ uncertainty for the experimental values, which is roughly the accuracy of the one loop calculation used here.
}
\label{tab:results}
\end{table}

\begin{table}[!h]
\centering 
\begin{tabular}{|rl|rl|r@{.}l|c|}
\hline
\multicolumn{2}{|c|}{Observable} & \multicolumn{2}{|c|}{Value at $m_{t}$} & \multicolumn{2}{|c|}{Best fit result} & Uncertainty\\
\hline
\hline
\rule{0 px}{15 px}
$\theta^\mns_{12}$ & in $^\circ$ & $33.57$&$ _{-0.75}^{+0.77}$ &\hspace{.55cm} $33$&$38$ &  $ _{-0.28}^{+0.30}$ \\
\rule{0 px}{15 px}
$\theta^\mns_{23}$ & in $^\circ$ & $41.4$&$_{-1.8}^{+3.5}$ & $39$&$06$ &  $ _{-1.32}^{+1.33}$ \\
\rule{0 px}{15 px}
$\theta^\mns_{13}$ & in $^\circ$& $8.75 $&$ _{-0.44}^{+0.42}$ & $9$&$41$ &  $ _{-0.27}^{+0.28}$ \\[2px]
\hline
\end{tabular}
\caption{Best fit results for the lepton mixing angles at $m_t(m_t)$, given here in degree for convenience.} \label{tab:results2}
\end{table}
\clearpage

In the following we turn to a discussion of the fit results and discuss how the model can be probed by present and future experiments. 
\begin{itemize}
\item
The sign of $\Delta m^2_\text{atm}$ will be in the focus of the next round of neutrino oscillation experiments and will allow to distinguish between inverse and normal neutrino mass ordering. 
Furthermore, in contrast to NH models our model predicts a sizeable effective mass for neutrino-less double beta decay experiments of
\begin{align}
m_{\beta\beta} & =\Big\vert (U_{e1}^\mns)^2 \sqrt{-\Delta m_\text{atm}^2 - 
      \Delta m_\text{sol}^2} + 
  (U_{e2}^\mns)^2 \sqrt{-\Delta m_\text{atm}^2}\Big\vert \nonumber \\
  & = (1.83_{-0.06}^{+0.05})\cdot 10^{-2} \, \eV\;,
\end{align}
in reach of future experiments \cite{Bilenky:2012qi}.

\item
The predictions for the PMNS phases are $\delta^\mns=180^\circ$ and $\varphi^\mns=180^\circ$. A measurement of $\delta^\mns$ would thus allow to test our model. Note that because $Y_\nu$ is real and the phase of the 2-1 element of $Y_e$ can be absorbed by the right-handed electron field, the PMNS phases are fixed and also do not run.

\item
The model features the relation\footnote{In fact our model realises $\theta_{12}^e \gtrsim \theta_C$ which corrects $\theta_{13}^\mns$ to larger values, as discussed in~\cite{Antusch:2012fb}.} $\theta_{13}^\mns \simeq \theta_C \sin\theta_{23}^\mns$, which is seen in the correlation found in the MCMC analysis as shown in Figure~\ref{subfig:t23t13}. While the result $\theta^\mns_{23}={39.06^\circ}_{-1.32^\circ}^{+1.33^\circ}$ is somewhat smaller than the currently reported experimental value in \cite{GonzalezGarcia:2012sz,Fogli:2012ua}, the result $\theta^\mns_{13} = {9.41^\circ}_{-0.27^\circ}^{+0.28^\circ}$ is slightly larger. Future, more precise measurements of the atmospheric and reactor angle will thus have the capability to falsify our model or to strengthen the relation $\theta_{13}^\mns \simeq \theta_C \sin\theta_{23}^\mns$ with higher precision.

\item
Due to the lepton mixing sum rule $\theta_{12}^\mns = \theta_{12}^\nu + \theta_{13}^\mns\cot\theta_{23}^\mns\cos\delta^\mns$ and $\delta^\mns=180^\circ$,
the angles $\theta_{13}^\mns$ and $\theta_{23}^\mns$ lead to a large deviation of  $\theta_{12}^\mns$ from $\theta_{23}^\nu=45^\circ$. We find $\theta^\mns_{12} = {33.38^\circ}_{-0.28^\circ}^{+0.30^\circ}$, which lies within $1\sigma$ from the experimental values as reported in \cite{GonzalezGarcia:2012sz}. The correlations shown in Figures~\ref{subfig:t12t13} and~\ref{subfig:t23t12} visualise the relations between $\theta_{12}^\mns$ and $\theta^\mns_{13}$, respectively $\theta_{12}^\mns$ and $\theta^\mns_{23}$.

\item
In the quark sector we predict the ratio $m_s/m_d= 18.55_{-0.24}^{+0.26}$, which follows from the specific set of Clebsch-Gordan coefficients used in the model and is not tuned by a parameter in the fit. Our result agrees well with $m_s/m_d = 18.9 \pm 0.8$, as obtained from experiments in Ref.~\cite{Leutwyler:1996qg}. A future more accurate determination of $m_s/m_d$ will provide an additional test of our model.
\end{itemize}

\subsection{Phenomenological differences to our recent normal hierarchy model}

Let us conclude this section with a comparison between the results of our inverse neutrino hierarchy model and the model in Ref.~\cite{nh_model}, which had a normal mass hierarchy.  Related to the negative sign of $\Delta m^2_\text{atm}$ the effective mass parameter of neutrino-less double beta decay $m_{\beta\beta}$ is predicted to $m_{\beta\beta}^{\text{(IH)}}=(1.83_{-0.06}^{+0.05})\cdot 10^{-2} \, \eV$ in reach of experiments and much larger than $m_{\beta\beta}^{\text{(NH)}}=(2.31_{-0.09}^{+0.12})\cdot 10^{-3} \, \eV$ in the NH model. 

The large value of $\theta_{13}^\mns$ opens up the possibility of measuring the Dirac CP phase $\delta^\mns$ in long-baseline neutrino oscillation experiments. Therefore the inverse- and normal mass hierarchy models can also clearly be distinguished via their respective predictions $\delta^\mns= 180^\circ\;\text{(IH)}$ and $\delta^\mns={268.79^\circ}_{-1.72^\circ}^{+1.32^\circ}\;\text{(NH)}$. The differences between the lepton mixing angles in the inverse- and normal mass hierarchy models turn out to be less than one degree. While we do not expect that new more precise measurements of $\theta_{13}^\mns$ and $\theta_{23}^\mns$ will allow to distinguish between the IH and NH predictions, $\theta^\mns_{12}$ could be measured in a future $\sim$60 km baseline reactor experiment with high precision~\cite{Minakata:2004jt}. This would allow to discriminate between the  predictions $\theta^\mns_{12}={33.38^\circ}_{-0.28^\circ}^{+0.30^\circ}\;\text{(IH)}$ and $\theta^\mns_{12}={34.29^\circ}_{-0.39^\circ}^{+0.35^\circ}\;\text{(NH)}$ of the two models. Note however that the fit results for $\theta^\mns_{12}$ and $\theta^\mns_{13}$ depend also on the experimental range for $\theta_{23}^\mns$. 

The models can furthermore be distinguished in the quark sector by their predictions for the CKM phase of $\delta^\ckm={69.27^\circ}_{-0.69^\circ}^{+0.91^\circ}\;\text{(IH)}$ and $\delta^\ckm={65.65^\circ}_{-0.53^\circ}^{+1.78^\circ}\;\text{(NH)}$, respectively.

\section{Summary}

We proposed an SU(5) GUT model with an $A_4$ family symmetry which leads to a strong inverse neutrino mass hierarchy.
In section 2 we discussed how such an inverse hierarchy can be obtained without fine-tuning, based on a novel vacuum alignment for $Y_\nu$ and an off-diagonal $M_R$. The model is constructed in a similar way as the NH model in Ref.~\cite{nh_model}, which allows a direct comparison. In particular, in the presented model we also realise the specific relation $\theta_{13}^\mns \simeq  \theta_C / \sqrt{2}$ from charged lepton mixing effects, linked to the quark sector mixing by GUT relations. Furthermore, CP-symmetry is also spontaneously broken by the VEVs of flavon fields.

On the other hand, the model has some characteristics which differ substantially from the NH model~\cite{nh_model}:  In addition to the inverse mass ordering, which makes the Majorana nature of neutrino masses testable by future $0\nu\beta\beta$ experiments, the presented model has a different mechanism to generate the parameters of the PMNS mixing matrix.
The angles arise from a maximal 1-2 mixing and zero 1-3 mixing in $m_\nu$ which are both modified by a charged lepton 1-2 mixing contribution that is linked to the Cabibbo angle $\theta_C$ via GUT relations.
On the other hand, the 2-3 mixing in $m_\nu$ can a priori take arbitrary values, so that~-- in contrast to the NH model~\cite{nh_model}~-- no charged lepton mixing effects are needed in order to reproduce the experimental tendency towards a non-maximal $\theta_{23}^\mns$.
In our model, the charged lepton 1-2 mixing has two important effects: (i) it induces $\theta_{13}^\mns \simeq  \theta_C / \sqrt{2}$ and (ii) it lowers $\theta_{12}^\mns$ significantly. 
The latter effect can bring $\theta_{12}^\mns$ in excellent agreement with the present measurement, as long as $\theta_{23}^\mns$ is significantly below $45^\circ$, as one can see from Eqs.~(\ref{eq:t13fromt12e}) and (\ref{eq:sumrule}). 
Finally, the IH model predicts that $\delta^\mns$ and $\varphi^\mns =\varphi_2^\mns - \varphi_1^\mns$ both are $180^\circ$, due to the specific phases of the flavon VEVs. 

In the phenomenological part of the paper, we confronted the model with the experimental data by performing a detailed fit including a Markov Chain Monte Carlo analysis for calculating the highest posterior density intervals (corresponding to the $1\sigma$ regions) for the model parameters and the observables. With 14 parameters and 18 measured observables, we obtain an excellent best-fit with $\chi^2/\textrm{d.o.f.}=1.1$, based on the present experimental data, which is even better than the fit of the NH model~\cite{nh_model}.
In addition, we also predict the values of the two currently unmeasured leptonic CP phases so that, with altogether 20 observables, the model makes 6 predictions. 
The results are presented in Tables~\ref{tab:pararesults}, \ref{tab:results} and~\ref{tab:results2}. 
We finally pointed out the differences between the predictions of the NH and IH model and discussed how the models can be discriminated with the data of running and future experiments.

\section*{Acknowledgements}
This project was supported by the Swiss National Science Foundation (projects 200021-137513 and CRSII2-141939-1).

\section*{Appendix}
\begin{appendix}

\section{The messenger sector}
As we pointed out in section~\ref{sec:model}, the construction of the messenger sector is of importance to assure the consistency of the effective superpotential. In Table~\ref{tab:A4messengerSector}, half of the messenger fields and their charges under gauge-, family-, shaping- and R-symmetry are listed. To each messenger field $\Phi_i$ a messenger $\bar\Phi_i$ exists, such that a mass term of the form $\Lambda_i \Phi_i \bar\Phi_i$ exists in the renormalizable superpotential. The fields $\bar\Phi_i$ are not listed in Table~\ref{tab:A4messengerSector}, since their charges can be trivially determined from the charges of the corresponding field $\Phi_i$.

In the Figures on pages \pageref{fig:diag1}ff we explicitly list the supergraph structure that determines how the operators of the effective superpotential in Eq.~(\ref{eq:superpotential}) are generated by integrating out the messenger fields.

\end{appendix}


\begin{table}[p]
\centering \small
\begin{tabular}{|c||cc|cc|cccc|cc|c|}
\hline
\rule{0 px}{15 px}
& SU(5) & $A_4$ & $\mathbb{Z}_2^{(a)}$ & $\mathbb{Z}_2^{(b)}$ & $\mathbb{Z}_6^{(a)}$ & $\mathbb{Z}_6^{(b)}$ & $\mathbb{Z}_6^{(c)}$ & $\mathbb{Z}_6^{(d)}$ & $U(1)_a$ & $U(1)_b$ & $U(1)_R$ \\[2ex] 
\hline\hline
\multicolumn{12}{|l|}{\rule{0 px}{15 px}Matter Fields} \\[1ex]
\hline\hline
\rule{0 px}{15 px}
$F$   & 	$\mathbf{\overline{5}}$  &$\mathbf{3}$ 	& \nl & \nl & 2 & \nl & \nl & 1 & 2 & \nl & 1	\\ 
$T_1$ & 	$\mathbf{10}$ &		\nlone & 1 & \nl & \nl & \nl & 1 & \nl & 1 & \nl & 1 \\
$T_2$ & 	$\mathbf{10}$ 	&		\nlone 	& \nl & \nl & \nl & 1 & \nl & \nl & 1 & \nl & 1	\\
$T_3$ & 	$\mathbf{10}$ & 		\nlone 	& \nl & 1 & \nl & \nl & \nl & \nl & 1 & \nl & 1	\\
$N_1$ & 	\nlone & 		\nlone  	& \nl & \nl & 4 & \nl & \nl & 2 & \nl & \nl & 1 \\
$N_2$ & 	\nlone & 		\nlone  	& \nl & \nl & 4 & \nl & \nl & 4 & \nl & \nl & 1 \\[1ex]
\hline\hline
\multicolumn{12}{|l|}{\rule{0 px}{15 px}Higgs Fields} \\[1ex]
\hline\hline
\rule{0 px}{15 px}
$H_5$ & 			$\mathbf{5}$ & 			\nlone 	& \nl & \nl & \nl & \nl & \nl & \nl & -2 & \nl & \nl \\
$H_{\bar 5}$ & 		$\mathbf{\overline{5}}$ & 		\nlone 	& \nl & \nl & \nl & \nl & \nl & \nl & \nl & -1 & \nl\\
$H_{45}$ & 		$\mathbf{45}$ & 			\nlone	& 1 & \nl & \nl & \nl & 2 & \nl & \nl & 1 & 2 \\
$H_{\overline{45}}$ & 	$\mathbf{\overline{45}}$ & 	\nlone 	& 1 & \nl & \nl & \nl & 4 & \nl & \nl & -1 & \nl \\
$H_{24}$     & 		$\mathbf{24}$ & 			\nlone 	& \nl & \nl & 4 & \nl & \nl & 5 & -3 & 1 & \nl \\[1ex]   
\hline\hline
\multicolumn{12}{|l|}{\rule{0 px}{15 px}Flavon Fields} \\[1ex] 
\hline\hline
\rule{0 px}{15 px}
$\phi_1$     & 		\nlone & $\mathbf{3}$ 	& \nl & \nl & \nl & \nl & \nl & 3 & \nl & \nl & \nl \\
$\phi_2$     & 		\nlone & $\mathbf{3}$ 	& \nl & \nl & \nl & \nl & 1 & \nl & \nl & \nl & \nl \\
$\phi_3$     & 		\nlone & $\mathbf{3}$ 	& \nl & 1 & \nl & \nl & \nl & \nl & \nl & \nl & \nl\\
$\phi_{ab}$  & 		\nlone & $\mathbf{3}$ 	& \nl & \nl & \nl & 5 & \nl & \nl & \nl & \nl & \nl\\
$\phi_{bc}$  & 		\nlone & $\mathbf{3}$ 	& \nl & \nl & \nl & \nl & \nl & 1 & \nl & \nl & \nl\\
$\xi_{12}$  & 		\nlone & \nlone  	& 1 & \nl & \nl & 1 & 1 & \nl & \nl & \nl & \nl\\
$\xi_{23}$  & 		\nlone & \nlone  	& \nl & 1 & \nl & 5 & \nl & \nl & \nl & \nl & \nl\\
$\xi_{M}$  & 		\nlone & \nlone  	& \nl & \nl & 1 & \nl & \nl & \nl & \nl & \nl & \nl \\[1ex]
\hline\hline
\multicolumn{12}{|l|}{\rule{0 px}{15 px}Driving Fields} \\[1ex] 
\hline\hline
\rule{0 px}{15 px}
$S$ & \nlone & \nlone              	& \nl & \nl & \nl & \nl & \nl & \nl & \nl & \nl & 2  \\
$D_{ab}^\alpha$ & 	\nlone & \nlone 	& \nl & \nl & \nl & 2 & \nl & \nl & \nl & \nl & 2  \\
$D_{ab}^\beta$ & 	\nlone & \nlone 	& \nl & \nl & \nl & 3 & \nl & \nl & \nl & \nl & 2  \\
$D_{ab}^\gamma$ & 	\nlone & \nlone 	& \nl & \nl & \nl & 4 & \nl & \nl & \nl & \nl & 2  \\
$D_{bc}^\beta$ & 	\nlone & \nlone 	& \nl & \nl & \nl & \nl & \nl & 3 & \nl & \nl & 2  \\
$D_{bc}^\gamma$ & 	\nlone & \nlone 	& \nl & \nl & \nl & \nl & \nl & 2 & \nl & \nl & 2  \\
$O_{1;2}$, $O_{1;2}^\prime$ & 	\nlone & $\mathbf{1}$, $\mathbf{1'}$ & \nl & \nl & \nl & \nl & 5 & 3 & \nl & \nl & 2  \\
$O_{1;3}$, $O_{1;3}^\prime$ & 	\nlone & $\mathbf{1}$, $\mathbf{1'}$ & \nl & 1 & \nl & \nl & \nl & 3 & \nl & \nl & 2  \\
$O_{2;3}$, $O_{2;3}^\prime$ & 	\nlone & $\mathbf{1}$, $\mathbf{1'}$ & \nl & 1 & \nl & \nl & 5 & \nl & \nl & \nl & 2  \\[1ex]
 \hline
\end{tabular}
  \caption{The matter-, Higgs-, flavon- and driving fields. A dot indicates that the field is an invariant singlet under the corresponding symmetry. Note that the $U(1)$ symmetries will get explicitly broken to $\mathbb{Z}_n$ symmetries by the Higgs sector.}
  \label{tab:A4MatterflavSector} 
\end{table}
\begin{table}[p]
\centering \small
\begin{tabular}{|c||cc|cc|cccc|cc|c|}
\hline
\rule{0 px}{15 px}
& SU(5) & $A_4$ & $Z_2^{(a)}$ & $Z_2^{(b)}$ & $Z_6^{(a)}$ & $Z_6^{(b)}$ & $Z_6^{(c)}$ & $Z_6^{(d)}$ &$U(1)_a$ & $U(1)_b$ & $U(1)_R$   \\[2ex]
\hline
\hline
\rule{0 px}{15 px}
$\Phi_1$ & $\mathbf{\overline{5}}$ & \nlone & \nl & \nl & 2 & \nl & \nl & 4 & 2 & \nl & 1 \\
$\Phi_2$ & $\mathbf{\overline{5}}$ & \nlone & \nl & \nl & 2 & \nl & \nl & 2 & 2 & \nl & 1 \\
$\Phi_3$ & $\mathbf{\overline{5}}$ & \nlone & \nl & 1 & \nl & 1 & \nl & \nl & 2 & \nl & 2 \\
$\Phi_4$ & \nlone & $\mathbf{3}$ & \nl & \nl & \nl & 2 & \nl & \nl & \nl & \nl & 2 \\
$\Phi_5$ & \nlone & $\mathbf{3}$ & \nl & \nl & \nl & \nl & \nl & 4 & \nl & \nl & 2 \\
$\Phi_6$ & $\mathbf{\overline{45}}$ & \nlone & \nl & \nl & \nl & \nl & 1 & \nl & -1 & 1 & 1 \\
$\Phi_7$ & $\mathbf{5}$ & \nlone & \nl & \nl & 4 & \nl & 5 & 5 & -2 & \nl & 1 \\
$\Phi_8$ & $\mathbf{\overline{10}}$ & \nlone & \nl & \nl & 2 & 5 & \nl & 1 & 2 & -1 & 1 \\
$\Phi_9$ & $\mathbf{5}$ & $\mathbf{3}$ & \nl & \nl & \nl & 1 & \nl & \nl & \nl & 1 & 2 \\
$\Phi_{10}$ & $\mathbf{\overline{5}}$ & \nlone & \nl & 1 & \nl & \nl & \nl & \nl & -1 & 1 & 1 \\
$\Phi_{11}$ & $\mathbf{5}$ & \nlone & \nl & 1 & 4 & \nl & \nl & 5 & -2 & \nl & 1 \\
$\Phi_{12}$ & $\mathbf{\overline{10}}$ & 3 & \nl & \nl & \nl & \nl & \nl & \nl & -1 & \nl & 1 \\
$\Phi_{13}$ & \nlone & $\mathbf{1'}$ & \nl & \nl & \nl & 2 & \nl & \nl & \nl & \nl & 2 \\
$\Phi_{14}$ & \nlone & $\mathbf{1''}$ & \nl & \nl & \nl & 2 & \nl & \nl & \nl & \nl & 2 \\
$\Phi_{15}$ & \nlone & \nlone & \nl & \nl & \nl & \nl & \nl & 4 & \nl & \nl & 2 \\
$\Phi_{16}$ & \nlone & \nlone & \nl & \nl & \nl & 4 & 4 & \nl & \nl & \nl & 2 \\
$\Phi_{17}$ & \nlone & \nlone & 1 & \nl & \nl & 3 & 3 & \nl & \nl & \nl & \nl \\
$\Phi_{18}$ & \nlone & \nlone & \nl & \nl & 4 & \nl & \nl & \nl & \nl & \nl & 2 \\
$\Phi_{19}$ & \nlone & \nlone & \nl & \nl & 3 & \nl & \nl & \nl & \nl & \nl & \nl \\
$\Phi_{20}$ & \nlone & $\mathbf{3}$ & \nl & 1 & \nl & 3 & \nl & \nl & \nl & \nl & \nl \\
$\Phi_{21}$ & \nlone & \nlone & \nl & \nl & \nl & \nl & 4 & \nl & \nl & \nl & 2 \\
$\Phi_{22}$ & \nlone & \nlone & \nl & \nl & \nl & \nl & 2 & \nl & \nl & \nl & 2 \\
$\Phi_{23}$ & $\mathbf{10}$ & \nlone & 1 & \nl & \nl & \nl & 5 & \nl & 1 & \nl & 1 \\
$\Phi_{24}$ & \nlone & $\mathbf{3}$ & \nl & \nl & \nl & \nl & \nl & 2 & \nl & \nl & 2 \\
$\Phi_{25}$ & \nlone & \nlone & 1 & \nl & \nl & 5 & 5 & \nl & \nl & \nl & \nl \\
$\Phi_{26}$ & \nlone & \nlone & \nl & \nl & \nl & \nl & \nl & 2 & \nl & \nl & 1 \\[1ex]

\hline
\end{tabular}
  \caption{The messenger fields used to generate the effective operators of the model.} \label{tab:A4messengerSector}
\end{table}

\begin{figure}[p]
\begin{center}
\includegraphics[page=1]{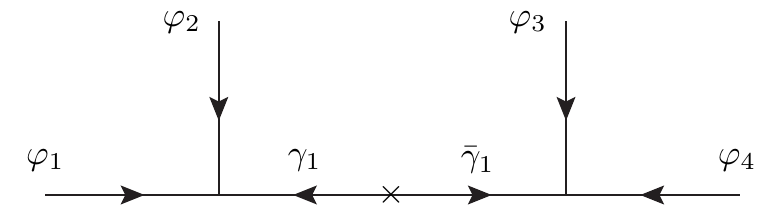}

\vspace{0.35cm}
\begin{tabular}{|c || c | c | c | c || c |}\hline\rule{0 px}{15 px}
operator&$\varphi_{1}$&$\varphi_{2}$&$\varphi_{3}$&$\varphi_{4}$&$\gamma_{1}$ \\[1ex]
\hline\rule{0 px}{15 px}
\#1 & $N_1$        & $H_{5}$     & $\phi_1$    & $F$            & $\Phi_1$ \\
\#2 & $N_2$        & $H_{5}$     & $\phi_{bc}$ & $F$            & $\Phi_2$ \\
\#3 & $\xi_{23}$   & $H_{5}$     & $T_2$       & $T_3$          & $\Phi_3$\\
\#4 & $\phi_{ab}$  & $\phi_{ab}$ & $\phi_{ab}$ & $D_{ab}^\beta$ & $\Phi_4$\\
\#5 & $\phi_{bc}$  & $\phi_{bc}$ & $\phi_{bc}$ & $D_{bc}^\beta$ & $\Phi_5$\\[1ex]
\hline
\end{tabular}
\caption{List of order 4 operators in the effective superpotential.} \label{fig:diag1}
\end{center}
\end{figure}

\begin{figure}
\begin{center}
\includegraphics[page=2]{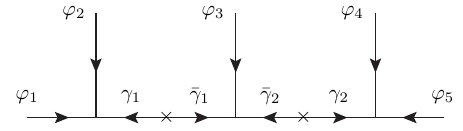}

\vspace{0.35cm}
\begin{tabular}{|c || c | c | c | c | c || c | c |} \hline\rule{0 px}{15 px}
operator&$\varphi_{1}$&$\varphi_{2}$&$\varphi_{3}$&$\varphi_{4}$&$\varphi_{5}$&$\gamma_{1}$&$\gamma_{2}$ \\[1ex]
\hline\rule{0 px}{15 px}
\#6  & $T_1$       & $H_{\overline{45}}$ & $H_{24}$        & $\phi_2$    & $F$          & $\Phi_6$ & $\Phi_7$\\
\#7  & $T_2$       & $H_{24}$            & $F$             & $\phi_{ab}$ & $H_{\bar 5}$ & $\Phi_8$   & $\Phi_9$\\
\#8  & $T_3$       & $H_{\bar 5}$        & $H_{24}$        & $\phi_3$    & $F$          & $\Phi_{10}$   & $\Phi_{11}$\\
\#9  & $T_2$       & $\phi_{ab}$         & $H_{5}$         & $\phi_{ab}$ & $T_2$        & $\Phi_{12}$   & $\Phi_{12}$  \\
\#10 & $\phi_{ab}$ & $\phi_{ab}$         & $D_{ab}^\gamma$ & $\phi_{ab}$ & $\phi_{ab}$  & $\Phi_4$   & $\Phi_4$  \\
\#11 & $\phi_{ab}$ & $\phi_{ab}$         & $D_{ab}^\gamma$ & $\phi_{ab}$ & $\phi_{ab}$  & $\Phi_{13}$  & $\Phi_{14}$   \\[1ex]
\#12 & $\phi_{bc}$ & $\phi_{bc}$         & $D_{bc}^\gamma$ & $\phi_{bc}$ & $\phi_{bc}$  & $\Phi_5$ & $\Phi_5$\\[1ex]
\#13 & $\phi_{bc}$ & $\phi_{bc}$         & $D_{bc}^\gamma$ & $\phi_{bc}$ & $\phi_{bc}$  & $\Phi_{15}$  & $\Phi_{15}$\\[1ex]
\hline
\end{tabular}
\caption{List of order 5 operators in the effective superpotential and corresponding messenger fields.} \label{fig:diag2}
\end{center}
\end{figure}

\begin{figure}
\begin{center}
\includegraphics[page=6]{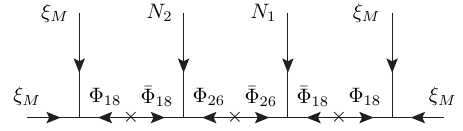}
\end{center}

\caption{The only effective operator of order 6 (\#14): $N_1 N_2 \xi_M^4$.} \label{fig:diag6}
\end{figure}

\begin{figure}
\begin{center}
\includegraphics[page=3]{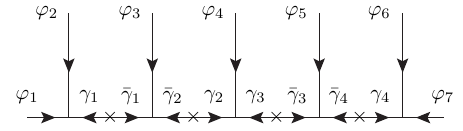}

\vspace{0.35cm}
\begin{tabular}{|c || c | c | c | c | c | c | c || c | c | c | c |} \hline
operator&$\varphi_{1}$&$\varphi_{2}$&$\varphi_{3}$&$\varphi_{4}$&$\varphi_{5}$&$\varphi_{6}$&$\varphi_{7}$&$\gamma_{1}$&$\gamma_{2}$&$\gamma_{3}$&$\gamma_{4}$ \\[1ex]
\hline\rule{0 px}{15 px}
\#15 & $\xi_{12}$  & $\xi_{12}$  & $\xi_{12}$ & $S$ & $\xi_{12}$ & $\xi_{12}$  & $\xi_{12}$  & $\Phi_{16}$ & $\Phi_{17}$  & $\Phi_{17}$  & $\Phi_{16}$ \\
\#16 & $\xi_M$     & $\xi_M$     & $\xi_M$    & $S$ & $\xi_M$    & $\xi_M$     & $\xi_M$     & $\Phi_{18}$ & $\Phi_{19}$  & $\Phi_{19}$  & $\Phi_{18}$ \\
\#17 & $\phi_{ab}$ & $\phi_{ab}$ & $\xi_{23}$ & $S$ & $\xi_{23}$ & $\phi_{ab}$ & $\phi_{ab}$ & $\Phi_4$  & $\Phi_{20}$ & $\Phi_{20}$ & $\Phi_4$  \\[1ex]
\hline
\end{tabular}
\caption{List of order 7 operators in the effective superpotential from supergraphs with linear topology.} \label{fig:diag3}
\end{center}
\end{figure}

\begin{figure}
\begin{center}
\includegraphics[page=4]{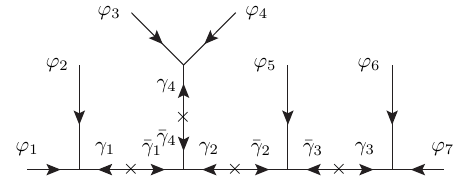}

\vspace{0.35cm}
\begin{tabular}{|c || c | c | c | c | c | c | c || c | c | c | c |} \hline\rule{0 px}{15 px}
operator&$\varphi_{1}$&$\varphi_{2}$&$\varphi_{3}$&$\varphi_{4}$&$\varphi_{5}$&$\varphi_{6}$&$\varphi_{7}$&$\gamma_{1}$&$\gamma_{2}$&$\gamma_{3}$&$\gamma_{4}$ \\[1ex]
\hline\rule{0 px}{15 px}
\#18 & $\phi_{2}$  & $\phi_{2}$  & $\phi_{2}$  & $\phi_{2}$  & $T_1$      & $T_1$       & $H_5$       & $\Phi_{21}$    & $\Phi_{22}$ & $\Phi_{23}$ & $\Phi_{21}$    \\
\#19 & $\phi_{bc}$ & $\phi_{bc}$ & $\phi_{bc}$ & $\phi_{bc}$ & $S$        & $\phi_{bc}$ & $\phi_{bc}$ & $\Phi_{15}$ & $\Phi_{24}$  & $\Phi_5$ & $\Phi_5$ \\
\#20 & $S$         & $\xi_{12}$  & $\xi_{12}$  & $\xi_{12}$  & $\xi_{12}$ & $\xi_{12}$  & $\xi_{12}$  & $\Phi_{25}$  & $\Phi_{17}$   & $\Phi_{16}$  & $\Phi_{16}$  \\
\#21 & $\phi_{2}$  & $\phi_{2}$  & $\phi_{2}$  & $\phi_{2}$  & $S$        & $\phi_{2}$  & $\phi_{2}$  & $\Phi_{21}$    & $\Phi_{22}$ & $\Phi_{21}$    & $\Phi_{21}$    \\[1ex]
\hline
\end{tabular}
\caption{List of order 7 operators in the effective superpotential from supergraphs with non-linear topology.} \label{fig:diag4}
\end{center}
\end{figure}

\begin{figure}
\begin{center}
\includegraphics[page=5]{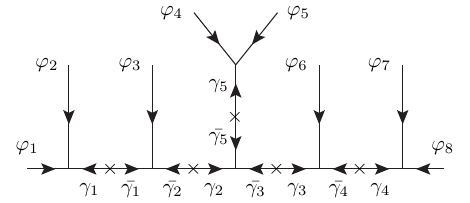}

\vspace{0.35cm}
\begin{tabular}{|c || c | c | c | c | c | c | c | c || c | c | c | c | c|} \hline\rule{0 px}{15 px}
operator&$\varphi_{1}$&$\varphi_{2}$&$\varphi_{3}$&$\varphi_{4}$&$\varphi_{5}$&$\varphi_{6}$&$\varphi_{7}$&$\varphi_{8}$&$\gamma_{1}$&$\gamma_{2}$&$\gamma_{3}$&$\gamma_{4}$&$\gamma_{5}$ \\[1ex]
\hline\rule{0 px}{15 px}
\#22 & $\xi_{12}$ & $\xi_{12}$ & $\xi_{12}$ & $\xi_{12}$  & $\xi_{12}$  & $T_2$  & $T_1$   & $H_5$   & $\Phi_{16}$  & $\Phi_{17}$   & $\Phi_{25}$      & $\Phi_{23}$ & $\Phi_{16}$ \\
\#23 & $\xi_M$    & $\xi_M$    & $N_1$      & $\phi_{bc}$ & $\phi_{bc}$ & $N_1$  & $\xi_M$ & $\xi_M$ & $\Phi_{18}$  & $\Phi_{26}$   & $\bar{\Phi}_{26}$ & $\Phi_{18}$  & $\Phi_{15}$ \\[1ex]
\hline
\end{tabular}
\caption{List of order 8 operators in the effective superpotential from supergraphs with non-linear topology.} \label{fig:diag5}
\end{center}
\end{figure}

\end{document}